\documentclass[aps,prb,reprint,superscriptaddress,amsmath,amssymb]{revtex4-2}
\preprint{APS/123-QED}
\usepackage{graphicx}
\usepackage{amsfonts}
\usepackage{appendix}
\usepackage[hidelinks]{hyperref}

\usepackage[usenames]{color} 

\begin{document}

\title{Three-Dimensional Domain-Wall Membranes}

\author{J. J. Mankenberg}
\affiliation{\mbox{Department of Physics and Electrical Engineering, Linnaeus University, SE-39231 Kalmar, Sweden}}
\email[]{jamaab@lnu.se}
\author{Ar. Abanov}
\affiliation{Department of Physics, Texas A\&M University, College Station, Texas 77843-4242, USA}

\date{\today}

\begin{abstract}
Three-dimensional magnetic textures, such as Hopfions, torons, and skyrmion tubes, possess rich geometric and topological structure, but their detailed energetics, deformation modes, and collective behavior are yet to be fully understood. In this work, we develop an effective geometric theory for general three-dimensional textures by representing them as embedded two-dimensional orientable domain-wall ``membranes". Using a local ansatz for the magnetization in terms of membrane coordinates, we integrate out the internal domain-wall profile to obtain a reduced two-dimensional energy functional. This functional captures the coupling between curvature, topology, and the interplay of micromagnetic energies, and is expressed in terms of a small set of soft-mode fields: the local wall thickness and in-plane magnetization angle. Additionally, we construct a local formula for the Hopf index which sheds light on the coupling between geometry and topology for nontrivial textures. We analyze the general properties of the theory and demonstrate its utility through the example of a flat membrane hosting a vortex as well as a toroidal Hopfion, obtaining analytic solutions for the wall thickness profile, associated energetics, and a confirmation of the Hopf index formula. The framework naturally extends to more complex geometries and can accommodate additional interactions such as Dzyaloshinskii–Moriya, Zeeman, and other anisotropies, making it a versatile tool for exploring the interplay between geometry, topology, and micromagnetics in three-dimensional spin systems.
\end{abstract}

\maketitle
\section{Introduction}\label{sec:intro}
Topologically nontrivial magnetic textures such as domain-walls, vortices, and skyrmions, have been extensively studied in two-dimensional systems, where their energetics, stability, dynamics, magnonics, and device applications are now relatively well understood \cite{Atkinson2003,Tretiakov2008,Vogel2023,Tomasello2017,Schwartz2022,Mller2017,Cherepov2012,Stepanov2017,Stepanova2022,Zhang2020,Zou2023,Blasing2020,Parkin2008,kosterlitz1973,Tomasello2014,Everschor-Sitte2018,Mermin1979,Papanicolaou1991}. In these systems, the relevant geometry is largely constrained to a plane, and the spatial variation of the magnetization can be parametrized in a way that simplifies both analytic treatments and numerical modeling. By contrast, the situation in three dimensions is fundamentally different: the texture can now bend, twist, and curve in ways that have no direct two-dimensional analog. This added geometric freedom means that the shape, curvature, and even topology of a three-dimensional texture are themselves dynamic degrees of freedom that can couple back to its micromagnetic energy, stability, and response to external perturbations.

In recent years, there has been significant progress in exploring genuinely three-dimensional magnetic configurations \cite{Donnelly2017,Fernndez-Pacheco2017,Donnelly2022,Fernndez-Pacheco2013,Ladak2022,May2021,Sanz-Hernndez2017,Sanz-Hernndez2018,Porrati2023,Raftrey2022,Fischer2020,Skoric2020,Girardi2024,Cheenikundil2023a,Cheenikundil2023b,Gu2022,Keller2018,Llandro2020,Winkler2019,Gubbiotti2025}. Experimental and numerical studies have reported the creation and observation of complex textures such as skyrmion strings \cite{Wolf2022,Seki2021,Birch2020,Schneider2018,Rybakov2013,Birch2022,Birch2021} and bobbers \cite{Rybakov2015,Charilaou2020,Mankenberg2025}, torons \cite{Mller2020,Zheng2018,Grelier2022,Raftrey2021,Liu2018,Li2022,Kuchkin2025}, Hopfions \cite{Wang2019,Rybakov2022a,Sutcliffe2018,Ishikawa1976,Radu2008,Ackerman2017,Tai2022,Chen2013,Tai2018,Voinescu2020,Liu2018,Raftrey2021,Kent2021}, and a multitude other field configurations characterized by interacting Bloch points and nontrivial linking of preimages \cite{Volkov2024,Yasin2024,Bassirian2022,Leonov2021,Rybakov2022b}. On the theoretical side, micromagnetic simulations and field-theoretical approaches have begun to address phenomena such as reconnection processes \cite{Li2022,Rybakov2015,Mankenberg2025} and defect-mediated transformations in three dimensions \cite{Vitelli2004,Birch2021,Mller2020,Schtte2014}. However, much of this work focuses on global properties such as topological charge or overall size, while the local geometric and structural details remain less systematically understood.

Although the existence and stability of three-dimensional magnetic textures are now firmly established, key questions remain unanswered regarding how their geometry, shape, and local structure influence their energetics and dynamics. These are not merely academic questions: in systems with strong Dzyaloshinskii–Moriya interaction, anisotropy, or magnetostatic effects, such geometric contributions can play a decisive role in determining stability thresholds, mode frequencies, and pathways for transformation or annihilation. In this work, we develop a general framework for resolving the energetics and dynamics of arbitrary three-dimensional textures into local contributions defined on an effective two-dimensional membrane corresponding to the texture’s core, or isosurface. This approach allows us to directly link local geometric properties, such as curvature and thickness, to measurable quantities, providing a systematic route toward understanding the interplay between geometry, topology, and micromagnetics in three dimensions.

\begin{figure}[ht]
\includegraphics[width=0.45\textwidth]{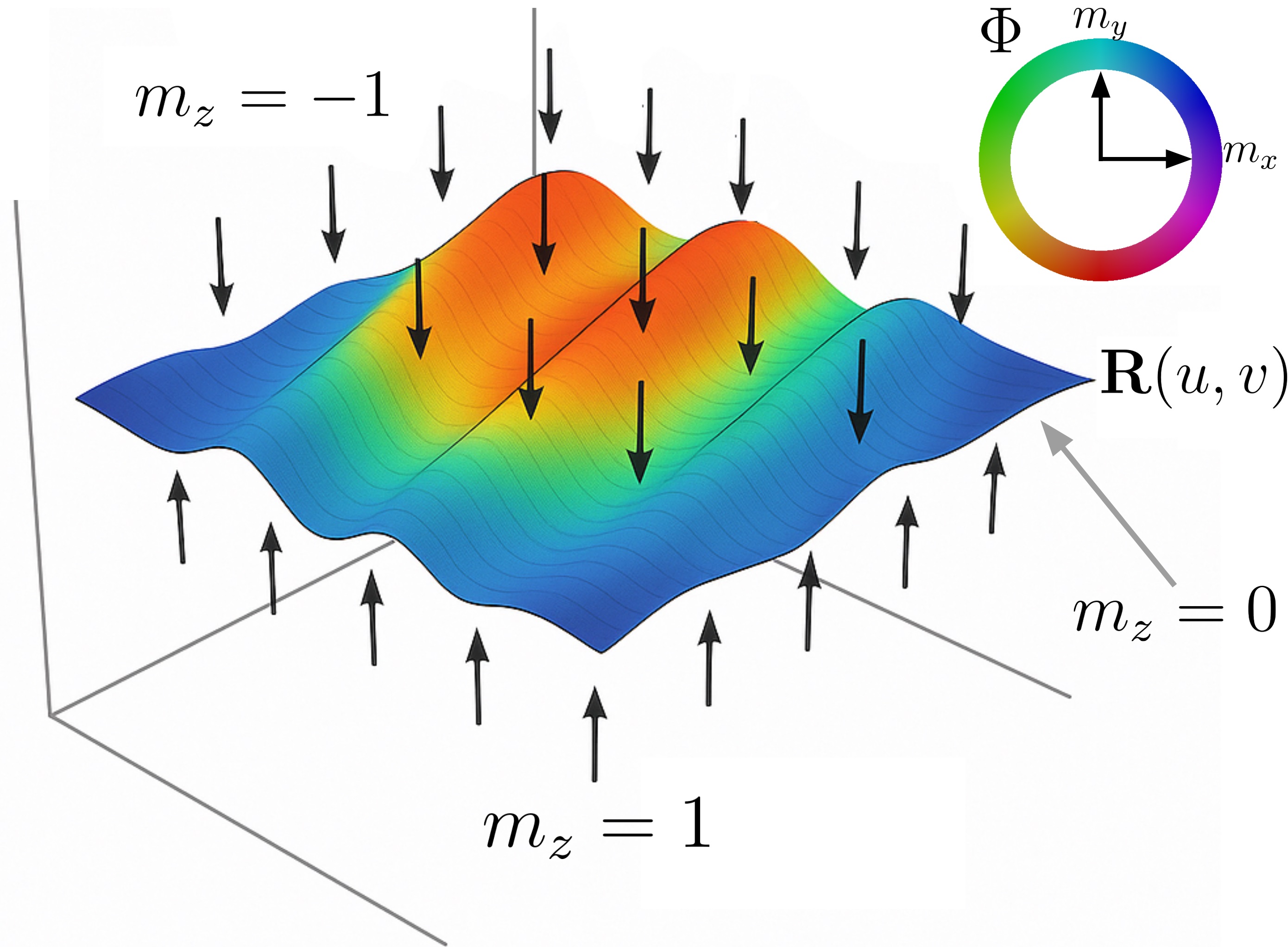} 
\caption{\label{fig:membrane1} Diagram of a smooth membrane separating domains of $m_z = 1$ and $m_z = -1$ defined as the surface $\mathbf{R}(u,v)$ in real space where $m_z = 0$, parametrized by the real numbers $u$ and $v$. On the membrane, the magnetization is reduced to the scalar field $\Phi(u,v)$ describing the spin angle in the $xy$-plane, here denoted by the color. The geometrical properties of a domain-wall membrane such as the local curvature and torsion are energetically coupled to the thickness and magnetization $\Delta(u,v)$ and $\Phi(u,v)$.}
\end{figure}

In Sec.~\ref{sec:two}, we present the definition and description of a domain-wall membrane as a two-dimensional geometric surface. We introduce the local magnetization, building from the flat-membrane case, and derive the corresponding local energy functional. In Sec.~\ref{sec:three}, we extend the framework to derive a local version of the equation of the Hopf invariant for compact membranes. In Sec.~\ref{sec:four}, we apply the general results of Sec.~\ref{sec:two} to the specific case of a flat membrane with a vortex and analyze its equilibrium properties, focusing on the interplay between wall width and winding number. In Sec.~\ref{sec:five}, we consider a toroidal Hopfion, confirming the local equation of the Hopf invariant and examine its energy landscape to explore its decay modes. Finally, in Sec.~\ref{sec:six} we summarize our conclusions and outline possible extensions of this geometric framework.

\section{Description and Energetics of Membranes}\label{sec:two}
We consider an arbitrarily large ferromagnetic sample with simple Heisenberg exchange and uniaxial anisotropy in the $z$-direction. The magnetization is described by the field $\mathbf{M}(\mathbf{r}) = M_s\mathbf{m}(\mathbf{r})$ for saturation magnetization $M_s$ and unit vector field $\mathbf{m}(\mathbf{r})$. The micromagnetic Hamiltonian is
\begin{equation}\label{eq:mmHamiltonian}
    \mathcal{H} = \int d^3r\bigg(\frac{J}{2}(\nabla\mathbf{m})^2 + \lambda(1-m_z^2)\bigg)
\end{equation}
where $J$ is the exchange constant and $\lambda$ is the anisotropy. We will limit our discussion here to this Hamiltonian because it allows for sufficient complexity in the space of possible structures. Future work will consider additional interactions such as Dzyaloshinskii-Moriya (DMI), Zeeman, as well as other types of anisotropy, however the structure and assumptions may change if, for example, the background is spiral rather than ferromagnetic. This Hamiltonian in Eq.~(\ref{eq:mmHamiltonian}) conventionally suggests a characteristic length scale, the fundamental domain-wall width $\Delta_0 = \sqrt{J/2\lambda}$ \cite{Hubert1998}, however as we will show in this work, structures smaller or larger than $\Delta_0$ not only exist but play an important role. Nonetheless, the anisotropy term favors a localization of deviations of $\mathbf{m}$ from the energy minima $(0,0,\pm1)$ with sizes on the order of $\Delta_0$. Thus, separating magnetic domains is a region of space where $m_z$ will change signs and therefore be zero.

We consider a domain-wall membrane in a three-dimensional sample to be an embedded two-dimensional surface
\begin{align}
    \mathbf{R}(u,v)
\end{align}
parametrized by the two real values $u$ and $v$. In this manner, the membrane is a smooth manifold in the classical sense, described by the mapping from some interval of $\mathbb{R}^2$ to $\mathbb{R}^3$. It is defined as the region of the sample separating domains where the magnetization is in the plane perpendicular to the anisotropy axis:
\begin{equation}
    m_z(\mathbf{R}) = 0.
\end{equation}
Thus, the magnetization on the membrane is fully described by a single angle $\Phi$ in the $xy$-plane as shown qualitatively in Fig.~\ref{fig:membrane1}.

We would like to emphasize that the membrane refers to a region of an otherwise arbitrary configuration of the magnetization embedded in a three-dimensional sample. It does not describe the geometry of the sample itself. The problems associated with curvature of the sample have been studied extensively in recent years by a number of groups\cite{Dao2025,Streubel2016,Gaididei2014,Turner2010,Makarov2022,Vitelli2004,Kamien2002,Sheka2022,Kravchuk2016}. Instead, it aligns more with previous work on two-dimensional theories\cite{Zhang2018,Rodrigues2018,Schroers2019a,Barton-Singer2020,Walton2020,Schroers2019b} and general geometric approaches applied to magnetic systems\cite{Hill2021,Dzyaloshinskii1978,Di2021,DiPietro2022,Chojnacki2024}.

Our goal is to decompose the energetics, topological properties, and relevant physical features of arbitrary three-dimensional magnetic textures into local structures defined on the membrane. In this paper, we tackle the equilibrium energetics. This approach is motivated by the fact that the nontrivial physics of such textures is concentrated in the vicinity of the membrane itself, where the magnetization varies strongly. Far from the membrane, the magnetization approaches a uniform configuration and can be regarded as energetically trivial. By focusing on the membrane, we reduce the full three-dimensional problem to a two-dimensional manifold embedded in three-dimensional space, on which all essential variations of the texture are localized. This reduction necessarily requires a geometric formulation of the problem that accounts for local curvature and its spatial variation since generic textures bend, twist, and deform in ways that cannot be captured by purely global parametrizations. The local geometric description enables us to express energies in terms of intrinsic and extrinsic curvature invariants, making the framework applicable to arbitrary configurations.

To initiate this construction, we must select a convenient local spin basis defined at each point along the membrane. This basis will serve as the bridge between the physical magnetization field and the geometric coordinates of the membrane. While many choices are available (each corresponding to a different gauge for representing the spin degrees of freedom) the one we will adopt here consists of the global unit vector $\hat{z}$, which couples spin-space to real space through the anisotropy, and two local unit vectors defined as
\begin{align}\label{eq:basis}
    \hat{\xi}(u,v) = \frac{\hat{z}\times\hat{n}(u,v)}{|\hat{z}\times\hat{n}(u,v)|}\text{ and } \hat{\chi}(u,v) = \hat{\xi}(u,v)\times\hat{n},
\end{align}
where $\hat{n}$ is the normal unit vector defined on the membrane as
\begin{equation}\label{eq:normal}
    \hat{n}(u,v) = \frac{\partial_u\mathbf{R}\times\partial_v\mathbf{R}}{|\partial_u\mathbf{R}\times\partial_v\mathbf{R}|}.
\end{equation}
This choice naturally incorporates the membrane’s geometry into the spin description, simplifies the projection of the micromagnetic fields and energies onto the membrane, and allows for a clear separation of in-plane and out-of-plane spin components. 

\begin{figure}[ht]
\includegraphics[width=0.45\textwidth]{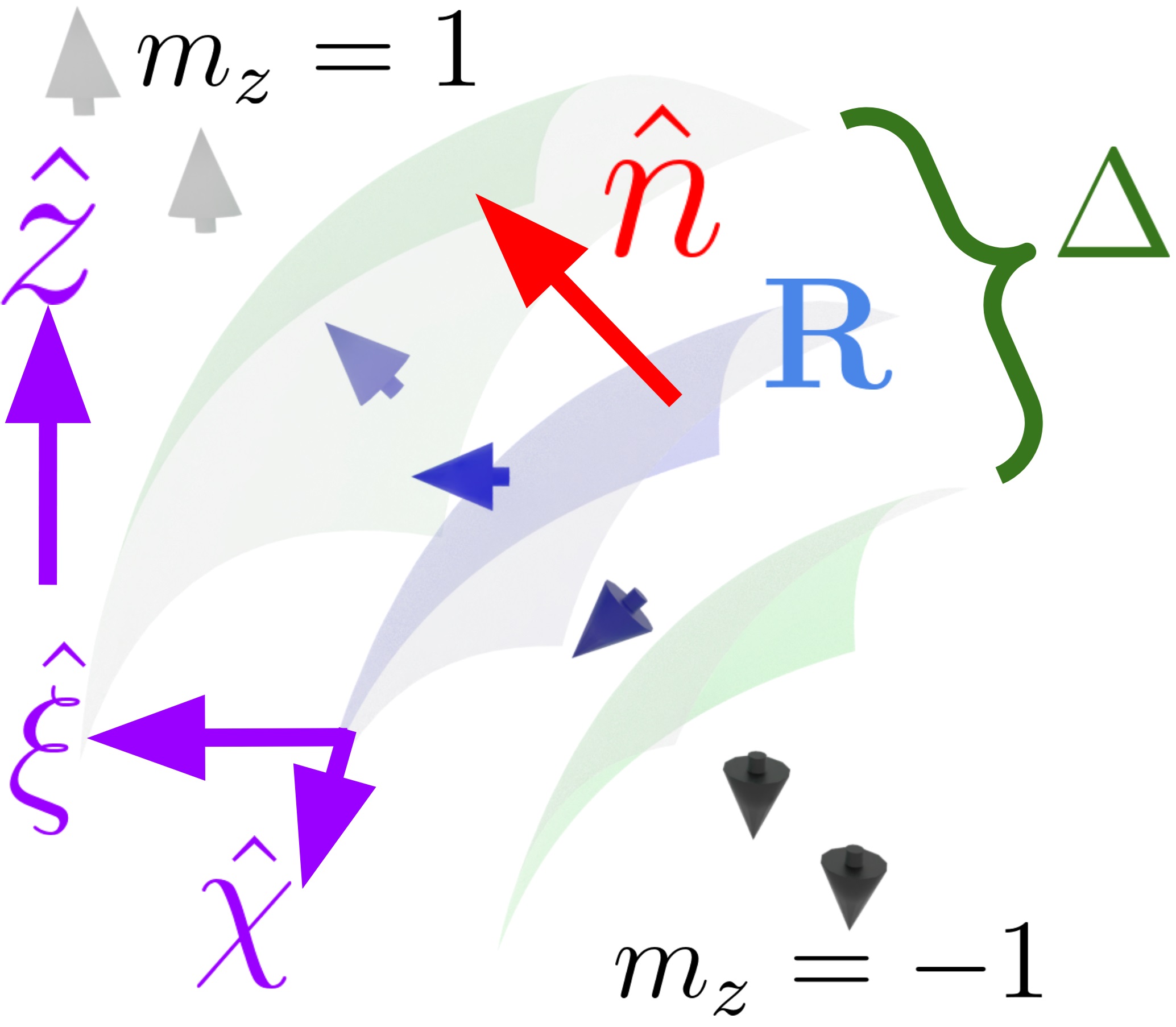}
\caption{\label{fig:membraneDetail} Schematic domain-wall structure across a curved membrane of thickness $\Delta$ separating domains of $m_z = 1$ and $m_z = -1$. On the membrane $\mathbf{R}$, the magnetization only has an in-plane component ($m_z = 0$). The spin basis convention used in this paper is also shown with the normal vector $\hat{n}$ to the membrane in red, the global background vector $\hat{z}$ in purple, and the two other local spin vectors $\hat{\xi}$ and $\hat{\chi}$ defined in Eq.~(\ref{eq:basis}) shown in purple.}
\end{figure}

The Hamiltonian in Eq.~(\ref{eq:mmHamiltonian}) can be expressed in the chosen local spin basis by introducing a magnetization ansatz that varies along the normal coordinate $n$ to the membrane surface. This ansatz describes the internal structure of the domain-wall that forms the membrane cross-section and is then integrated out to obtain an effective theory defined solely on the membrane. Concretely, the membrane geometry is specified by the surface embedding $\mathbf{R}(u,v)$, and the local normal $\hat{n}$ at each point defines a one-dimensional coordinate line along which the magnetization profile is taken to follow the form of a domain-wall (see Fig.~\ref{fig:membraneDetail} and Appendix \ref{app:appA}).

There is considerable freedom in the choice of the domain-wall type used for the cross-section: Bloch, N\'eel, intermediate/canted, head-to-head or tail-to-tail, twisted or precessional, chiral, or other variants that arise from specific micromagnetic interactions. In the present treatment, these different types are all solutions of the same one-dimensional Euler–Lagrange equations obtained from the micromagnetic energy functional for the uniform, flat case. Consequently, integrating over $n$ yields the same form of the effective membrane energy up to possible differences in numerical prefactors that depend on the specific wall profile \footnote{This equivalence holds provided the underlying Hamiltonian and boundary conditions are the same; see Appendix \ref{app:appA} for a detailed discussion.}. The general one-dimensional wall profile is therefore first obtained for the simplest case of a flat, uniform membrane and then used as the local cross-sectional solution for the fully curved membrane.

Substituting this flat-domain-wall solution into the micromagnetic energy functional and performing the integration over $n$ (see Appendix \ref{app:appB}), one arrives at the effective total energy and the first main result of this paper:
\begin{align}\label{eq:totalEnergy}
\mathcal{H}[\Delta, \Phi, \mathbf{R}] = \int dudv \sqrt{\det g}~ \Sigma(\Delta, \Phi, \mathbf{R}),
\end{align}
where $g$ is the metric tensor associated with the membrane’s embedding $\mathbf{R}(u,v)$, and $\det g$ is the corresponding surface Jacobian determinant (the area element in $(u,v)$-coordinates). The quantity $\Sigma$ is the effective energy density on the membrane,
\begin{align}\label{eq:EnDensity}
\Sigma = \sigma(\Delta) + \frac{J}{2} \left[ \Delta_\alpha \Delta_\beta\Pi^{\Delta}_{\alpha\beta} + \Phi_\alpha \Phi_\beta \Pi^{\Phi}_{\alpha\beta} \right],
\end{align}
where
\begin{equation}\label{eq:surfTens}
\sigma(\Delta) = J\left(\frac{1}{\Delta}+\frac{\Delta}{\Delta_0^2}\right),
\end{equation}
is the “surface-tension” term whose minimum $2J/\Delta_0$ is at $\Delta=\Delta_0$, the fundamental wall thickness.

Here $\Delta(u,v)$ and $\Phi(u,v)$ are the slowly varying collective coordinates describing, respectively, the local width of the domain-wall and the in-plane spin angle. The antisymmetric (dual) derivatives are defined by
\begin{align}
\Delta_\alpha = \epsilon_{\alpha\beta}\partial_\beta \Delta, \text{ and }
\Phi_\alpha = \epsilon_{\alpha\beta}\partial_\beta \Phi,
\end{align}
where $\epsilon_{\alpha\beta}$ is the Levi–Civita symbol on the surface.

The geometric coupling tensors $\Pi^{\Delta}_{\alpha\beta}$ and $\Pi^{\Phi}_{\alpha\beta}$ are elliptic operators that encode how gradients of $\Delta$ and $\Phi$ couple to the surface geometry; their explicit forms are given in Appendix \ref{app:appB}. The first term in Eq.~(\ref{eq:EnDensity}) is proportional to the membrane area (once multiplied by $\det g$), and thus plays the role of a nonuniform surface tension whose strength depends on the local wall width.

\begin{figure}[ht]
\includegraphics[width=0.45\textwidth]{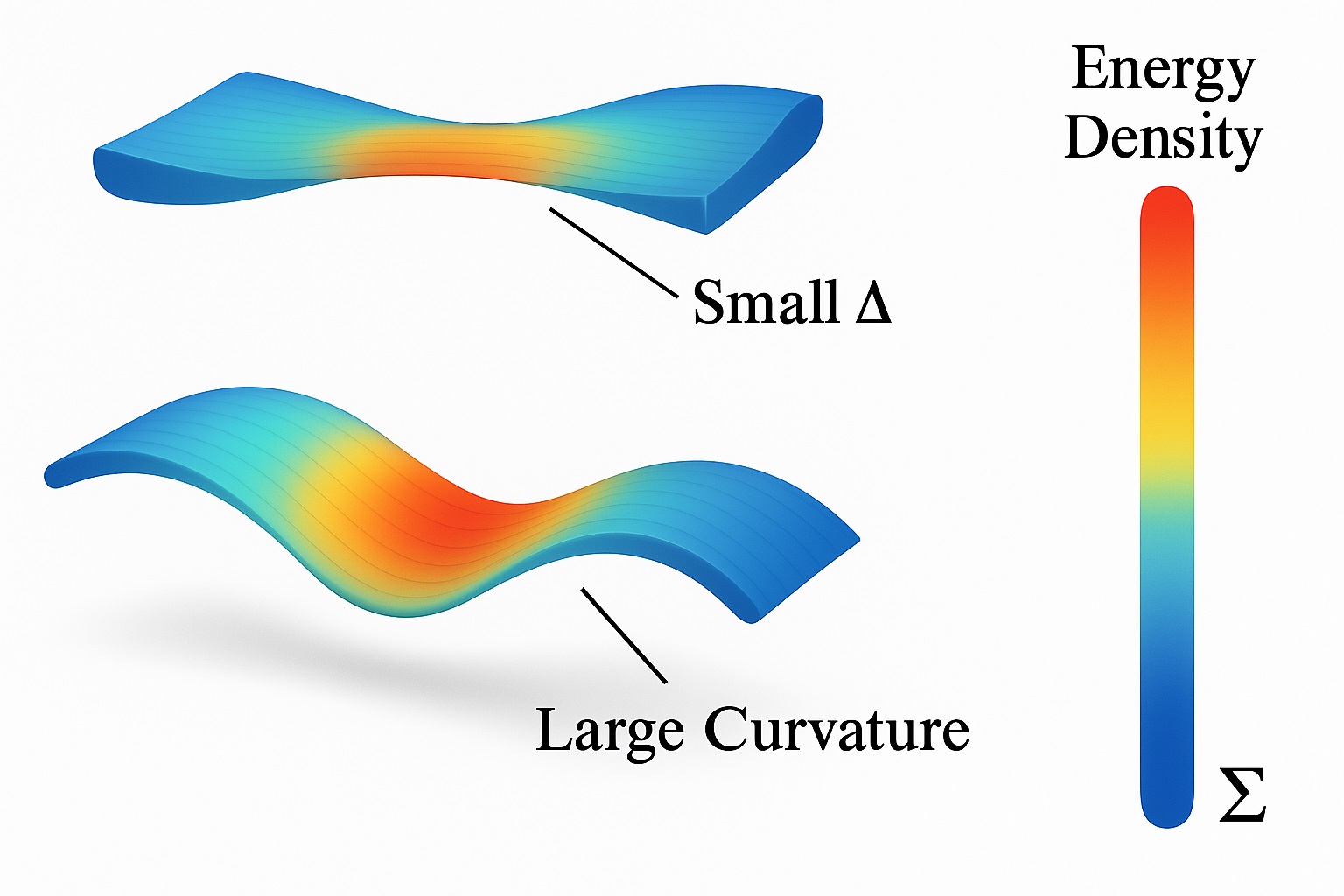} 
\caption{\label{fig:membraneDensity} Qualitative illustration of geometry-induced energy variations in the membrane. The top panel shows that reducing the local thickness $\Delta$ enhances the energy density, while the bottom panel demonstrates that increased curvature likewise raises the energy cost. The color scale indicates the local energy density $\Sigma$ from Eq.~(\ref{eq:EnDensity})}
\end{figure}

Physically, the structure of Eq.~(\ref{eq:EnDensity}) is analogous to that of an elastic surface with additional internal degrees of freedom. The term $\sigma(\Delta)$ acts as an effective surface tension: reducing the wall width $\Delta$ increases exchange energy, while increasing it incurs anisotropy cost, producing the characteristic form. The gradient terms involving $\Pi^{\Delta}_{\alpha\beta}$ and $\Pi^{\Phi}_{\alpha\beta}$ represent geometric stiffnesses: penalties for spatial variation of the wall width $\Delta$ or the internal spin rotation $\Phi$ along the membrane surface. The antisymmetric derivatives $\Delta_\alpha$ and $\Phi_\alpha$ naturally arise from the local spin-basis choice and encode the fact that variations of $\Delta$ and $\Phi$ couple to the membrane geometry. Together, these contributions define a two-dimensional effective field theory in which the membrane shape $\mathbf{R}(u,v)$ and the internal fields $\Delta(u,v)$ and $\Phi(u,v)$ are coupled through the metric $g_{\alpha\beta}$ and the curvature-dependent operators $\Pi^{\Delta}_{\alpha\beta}$ and $\Pi^{\Phi}_{\alpha\beta}$.

This effective description captures the full three–dimensional micromagnetic energetics of a complex texture in terms of a reduced set of geometric and internal fields defined on a two–dimensional manifold. By integrating out the the strong variation of the magnetization across the membrane, the model isolates the soft, collective degrees of freedom that dominate the large–scale behavior. This makes it possible to analyze how membrane geometry, through its local curvature and metric, directly influences the energetics and dynamics of the underlying magnetic texture (see Fig.~\ref{fig:membraneDensity} for a qualitative diagram). It also establishes a natural bridge between micromagnetics and theories of elastic surfaces, enabling the use of well–developed tools from differential geometry and soft–matter physics to study magnetic systems \cite{LandauElasticity,Kamien2002}.

In practical terms, the resulting membrane theory provides a tractable framework for exploring a wide range of phenomena without resorting to full three–dimensional micromagnetic simulations. It can be used to predict how curvature and topology affect the stability of magnetic textures or to design membranes whose shape and internal spin structure are tailored for specific dynamical responses. For example, the coupling between $\Delta$, $\Phi$, and geometry could be exploited for mechanically tunable spintronics, where elastic deformations modify the propagation of spin waves. Because the model retains the essential energetics in a compact form, it also opens the door to analytical studies of stability, mode spectra, and topological properties that would be prohibitively difficult in the full three-dimensional problem.

In the present formulation, boundaries and edge effects have been omitted for clarity, with the analysis focused on membranes without physical terminations or interfaces. In realistic systems, however, such effects can play a decisive role, particularly in the presence of DMI, magnetostatic coupling, or spatially varying material parameters, all of which can induce boundary–localized twists or alter the stability of bulk modes. Extending the framework to include boundary contributions and appropriate edge conditions would therefore be an important addition, enabling the treatment of finite membranes, patterned geometries, and hybrid structures where edge–geometry coupling becomes a dominant factor.

\section{Hopf Index in Local Coordinates}\label{sec:three}
In this section, we address the Hopf invariant, the central topological invariant characterizing spin structures in three dimensions, within the framework of our local membrane formalism. Because the essential features of a spin texture are concentrated in the vicinity of the membrane, the relevant topological information is likewise encoded there. This allows us to reduce the evaluation of the Hopf invariant to an integral defined solely over the membrane.

The standard expression for the Hopf invariant used in micromagnetics, often referred to as the Whitehead formula \cite{Whitehead1947}, is  
\begin{equation}\label{eq:hopf}
    H = \frac{-1}{(2\pi)^2}\int dV~\mathbf{F}\cdot\mathbf{A},
\end{equation}  
where  
\begin{equation}
    F_i = \frac{1}{2}\epsilon_{ijk}\,\mathbf{m}\cdot(\partial_j\mathbf{m}\times\partial_k\mathbf{m})
\end{equation}  
defines the so-called emergent magnetic field, and $\mathbf{A}$ is its associated vector potential, satisfying  
\begin{equation}
    \mathbf{F} = \nabla\times\mathbf{A}.
\end{equation}  
The Hopf invariant is conserved under uniform boundary conditions, i.e. when $\mathbf{m}\to \text{const}$ as $|\mathbf{r}|\to\infty$. This requires that the membrane under consideration be compact and sufficiently far from the system’s boundaries. While generalizations exist for more complicated boundary conditions \cite{Knapman2024,Azhar2025,Prior2014,Cantarella2001,Berger1993,Berger2006,Berger1984,Shtilman1985,Berger1985,Ryder1980}, we will not consider them here.  

By substituting the domain-wall ansatz into Eq.~(\ref{eq:hopf}) and working with two local patches for the vector potential (since it cannot be globally defined), we obtain the reduced expression  
\begin{align}\label{eq:localHopf}
    H = \frac{-1}{(2\pi)^2}\int dudv\,\mathcal{K},
\end{align}  
with  
\begin{align}
    \mathcal{K} = \begin{vmatrix}
    \partial_u\Phi & \partial_v\Phi \\
    \Gamma_{\chi\xi u} & \Gamma_{\chi\xi v}
\end{vmatrix}.
\end{align}  
Details of the derivation are provided in Appendix \ref{app:appC}. Here,  
\begin{align}
    \Gamma_{\chi\xi u} = \partial_u\hat{\xi}\cdot\hat{\chi}, \qquad
    \Gamma_{\chi\xi v} = \partial_v\hat{\xi}\cdot\hat{\chi}
\end{align}  
are Christoffel coefficients, and the determinant captures the difference between the variation of $\Phi$ in the $(u,v)$ directions and the intrinsic rotation of the local $\{\hat{\chi},\hat{\xi}\}$ frame. In two-dimensional geometry or in line-bundle (U(1)) contexts \cite{Schroers2019b,Rybakov2022b}, such differences between partial derivatives and connection coefficients naturally yield a local curvature.  

For compact membranes, the integration over $u$ and $v$ is periodic, ensuring that $H$ takes integer values. Because the curvature-like determinant $\mathcal{K}$ is a geometric quantity, Eq.~(\ref{eq:localHopf}) bears a striking resemblance to the Gauss-Bonnet theorem, which links geometric curvature to topological invariants.  

In Sec.~\ref{sec:five}, we will specialize to a toroidal membrane and show explicitly that, for a standard Hopfion configuration, Eq.~(\ref{eq:localHopf}) reproduces the expected Hopfion linking number.

\section{Example 1: Flat Membrane with a Vortex}\label{sec:four}
As a demonstration of a simple and insightful application, let us consider a flat domain-wall in the $xy$-plane with a vortex. In a sense, we can view this as the XY model embedded in three dimensions.

Recall that the metric tensor $g_{\alpha\beta} = \partial_\alpha\mathbf{R}\cdot\partial_\beta\mathbf{R}$ determines the local area element $\sqrt{\det g}dudv$. The geometric coupling tensors $\Pi^{\Delta}_{\alpha\beta}$ and $\Pi^{\Phi}_{\alpha\beta}$ in Eq.~(\ref{eq:EnDensity}) are obtained by integrating over the normal coordinate $n$ (see Appendix \ref{app:appB} for their exact forms) and are dependent on the membrane geometry through $g_{\alpha\beta}$.

Specializing to a flat membrane, we adopt plane–polar coordinates $$(u,v)\equiv(r,\phi)$$ as parameters, with embedding
\begin{align}
\mathbf{R} = \begin{pmatrix}
r\cos\phi \\
r\sin\phi \\
z_0
\end{pmatrix}.
\end{align}
This describes a flat plane at height $z_0$. In this case, the unit normal $\hat{n}$ is constant simplifying the expressions for the coupling tensors. They are, along with the metric and its determinant,
\begin{align}
g_{\alpha\beta} = \begin{pmatrix}
1 & 0 \\
0 & r^2
\end{pmatrix},~ \det g = r^2,~ \Pi^\Phi = \frac{2\Delta}{r^2} \begin{pmatrix}
1 & 0 \\
0 & r^2
\end{pmatrix}, \\ \text{ and } \Pi^\Delta = \frac{\pi^2}{6} \frac{1}{\Delta r^2} \begin{pmatrix}
1 & 0 \\
0 & r^2
\end{pmatrix}.
\end{align}

\begin{figure}[ht]
\includegraphics[width=0.45\textwidth]{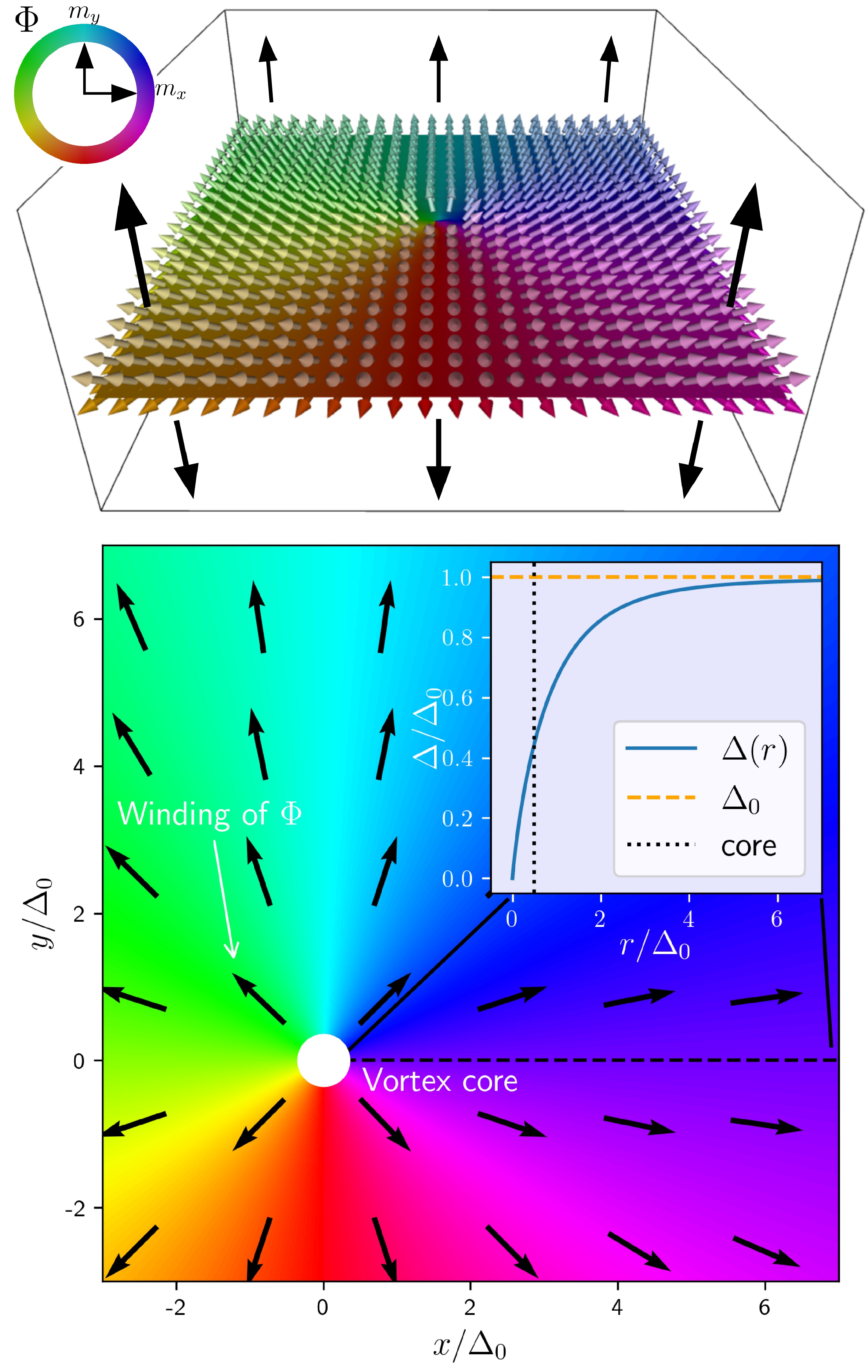} 
\caption{\label{fig:flat_vortex} A flat planar membrane with a vortex with singular winding such as that considered in Sec.~\ref{sec:three}. This membrane is parametrized by $(u,v)\equiv(r,\phi)$ in plane-polar coordinates. The effective energy functional derived in Eq.~(\ref{eq:totalEnergy}) is minimized by $\Delta(r)$. The solution is plotted in the inset revealing how the addition of a topological whirl can affect the structure of the membrane.  The "core" region denotes the limit of the small $r$ solution found in Eq.~(\ref{eq:DeltaSmallr}). An effective chemical potential is found for the Bloch point in Eq.~(\ref{eq:chemPot}).}
\end{figure}

We can assume that the domain-wall thickness $\Delta$ and in-plane angle are circularly symmetric, so take $\Delta$ to depend only on $r$, and $\Phi = n\phi$ meaning there is an integer $n$ winding:
\begin{align}
\Delta_\alpha = \begin{pmatrix}
0 \\
-\partial_r \Delta
\end{pmatrix}, \text{ and } \Phi_\alpha = \begin{pmatrix}
n \\
0
\end{pmatrix}.
\end{align}\label{eq:EDelta}

Substituting these quantities into the energy in Eq.~(\ref{eq:totalEnergy}) gives
\begin{align}
E[\Delta(r)] = 2\pi \int \left[r\sigma(\Delta) + \frac{J\pi^2}{12} \frac{r}{\Delta} (\partial_r \Delta)^2 + \frac{Jn^2 \Delta}{r} \right] dr.
\end{align}

$\Delta(r)$ is the unknown function and will minimize this energy functional according to its corresponding Euler-Lagrange equation:
\begin{equation}\label{eq:vortexEL}
r \frac{\partial \sigma}{\partial \Delta} - \frac{J\pi^2}{12} \frac{1}{r} \left( r \frac{\partial_r \Delta}{\Delta} \right)^2 - \frac{J\pi^2}{6} \partial_r \left( r \frac{\partial_r \Delta}{\Delta} \right) + \frac{Jn^2}{r} = 0.
\end{equation}

Note that if $n = 0$, then $\Delta = \Delta_0$ since this is just the trivial flat membrane.  If $n \neq 0$, for $r \to \infty$ the solution should have a form $\Delta(r) = \Delta_0 + \delta(r)$. If we now linearize Eq.~(\ref{eq:vortexEL}) with respect to $\delta$ and find the solution with $\delta(r \to \infty) \to 0$ we get the relation
\begin{equation}
\delta(r) = - \frac{Jn^2}{\partial^2 \sigma / \partial \Delta^2_0} \frac{1}{r^2}.
\end{equation}
Using Eq.~(\ref{eq:surfTens}) for the surface tension $\sigma$,
\begin{equation}\label{eq:DeltaLarger}
\Delta(r \to \infty) \approx \Delta_0\left(1 -  \frac{n^2 \Delta_0^2}{4} \frac{1}{r^2}\right).
\end{equation}
This asymptotic form is valid in the regime $r \gg |n|\Delta_0/2$, indicating that the characteristic size of the vortex core is set by $\Delta_0$ as well as the winding number. Physically, this means that far from the core the membrane thickness approaches $\Delta_0$, with only a small $1/r^2$ correction.

In the opposite limit as $r \to 0$, a polynomial solution approaches
\begin{equation}\label{eq:DeltaSmallr}
\Delta(r) = \frac{r}{\sqrt{n^2 - \pi^2 / 12}},
\end{equation}
so that the membrane thickness decreases linearly to zero at the Bloch point. We emphasize the universal character of the decrease of the membrane thickness approaching a Bloch point. Equation \eqref{eq:DeltaSmallr} shows that it depends only on the winding number and does not depend on any parameters. A core structure in which $\Delta$ vanishes linearly with $r$ is a feature that, to our knowledge, has not been reported previously. It is a direct outcome of the present formulation and offers a slightly different perspective on the structure.

We can examine some implications of these solutions by considering the difference in energy $\mathcal{E}$ between a flat featureless membrane and this one with the vortex. This quantity plays the role of the chemical potential associated with creating a Bloch point or vortex:
\begin{align}
    \mathcal{E} =& E[\Delta(r)] - E[\Delta_0] \\=& 2\pi \int dr \bigg[ r(\sigma(\Delta) - \sigma(\Delta_0)) \\&+ \frac{J\pi^2}{12}\frac{r}{\Delta}(\partial_r\Delta)^2 + \frac{Jn^2\Delta}{r}\bigg]
\end{align}
Here, $\Delta$ denotes the solution of Eq.~(\ref{eq:vortexEL}). The integral diverges logarithmically at large $r$ exactly the same way as if we used constant $\Delta_0$. For small $r$ in the case $\Delta=\Delta_0$ the energy also diverges logarithmically. This divergence typically is taken care of by introducing a cutoff at distances of the order of $\Delta_0$. However, if  we use $\Delta(r)$ from Eq.~(\ref{eq:DeltaSmallr}) we find that unlike the usual case the integral does not diverge. In fact, integrating up to the size of the core $a$ such that $\Delta(a)=\Delta_0$ we find the core energy difference to be
\begin{equation}\label{eq:chemPot}
    \mathcal{E}_{core} = 2\pi J\Delta_0\left(\frac{1}{3}\sqrt{n^2-\frac{\pi^2}{12}} + n^2+\frac{\pi^2}{12}\right).
\end{equation}

This simple case demonstrates how the geometric framework captures the interplay between curvature, topology, and micromagnetic energy even in the simplest configurations. Imposing a topologically protected vortex at the origin, we obtain an analytic solution that reveals how energy concentrates around the core while remaining finite and well-behaved at large distances. This example validates the membrane formalism in a limiting case and serves as a strong statement on the foundation for exploring more complex geometries where curvature and topology play an even more pronounced role.

\begin{figure}[ht]
\includegraphics[width=0.45\textwidth]{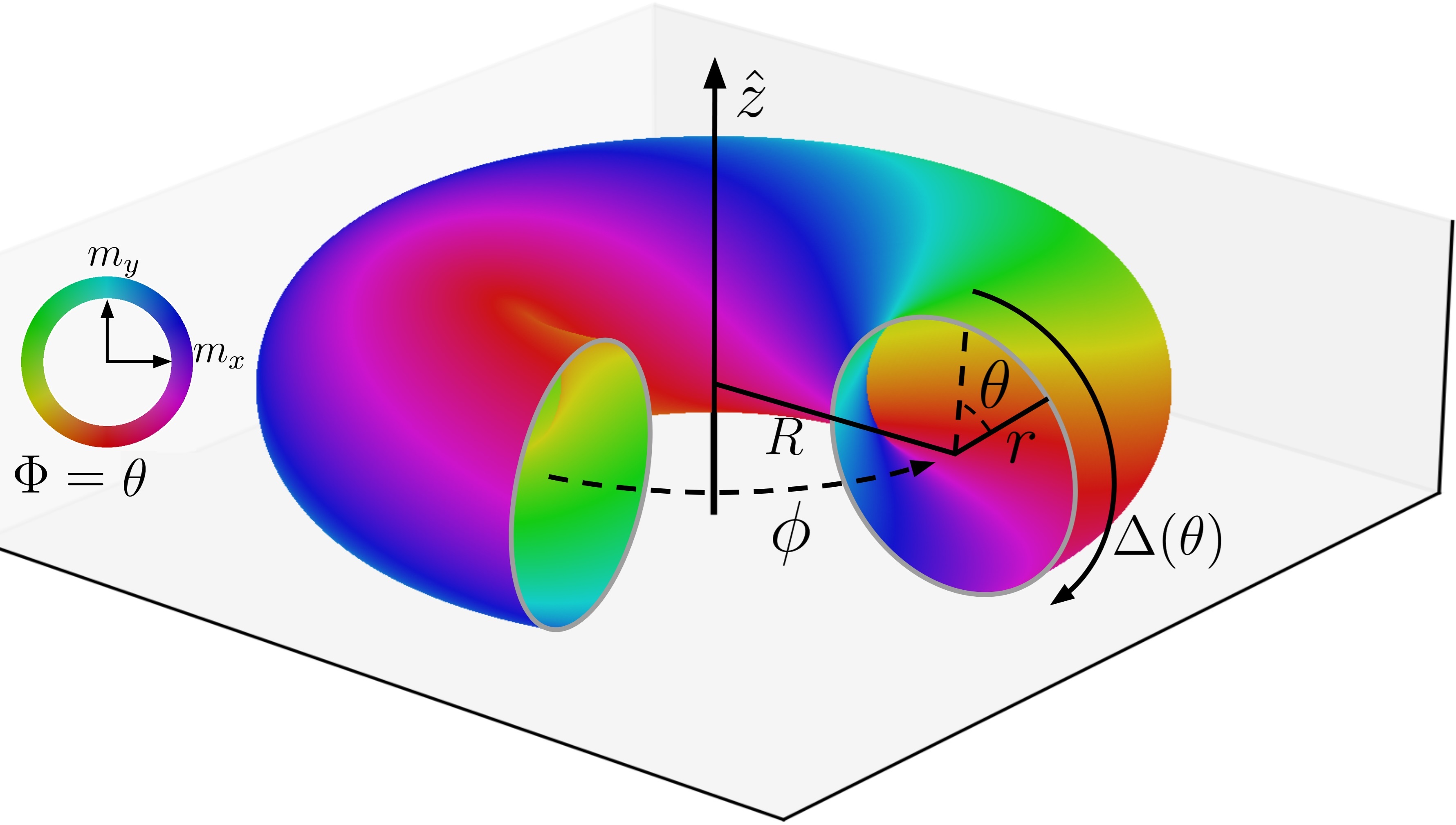} 
\caption{\label{fig:torus} A schematic of the toroidal membrane hosting a Hopfion of linking 1 considered in Sec.~\ref{sec:five}. The membrane is parametrized by the poloidal and toroidal angles $\theta$ and $\phi$ respectively which span $[0,2\pi)$ as shown in Eq.~(\ref{eq:torus}). The in-plane angle $\Phi = \theta$ gives a single Hopf linking and the domain-wall thickness $\Delta$ is taken to have only a $\theta$ dependence due to the symmetries of the system.}
\end{figure}

\begin{figure*}[ht]
\includegraphics[width=\textwidth]{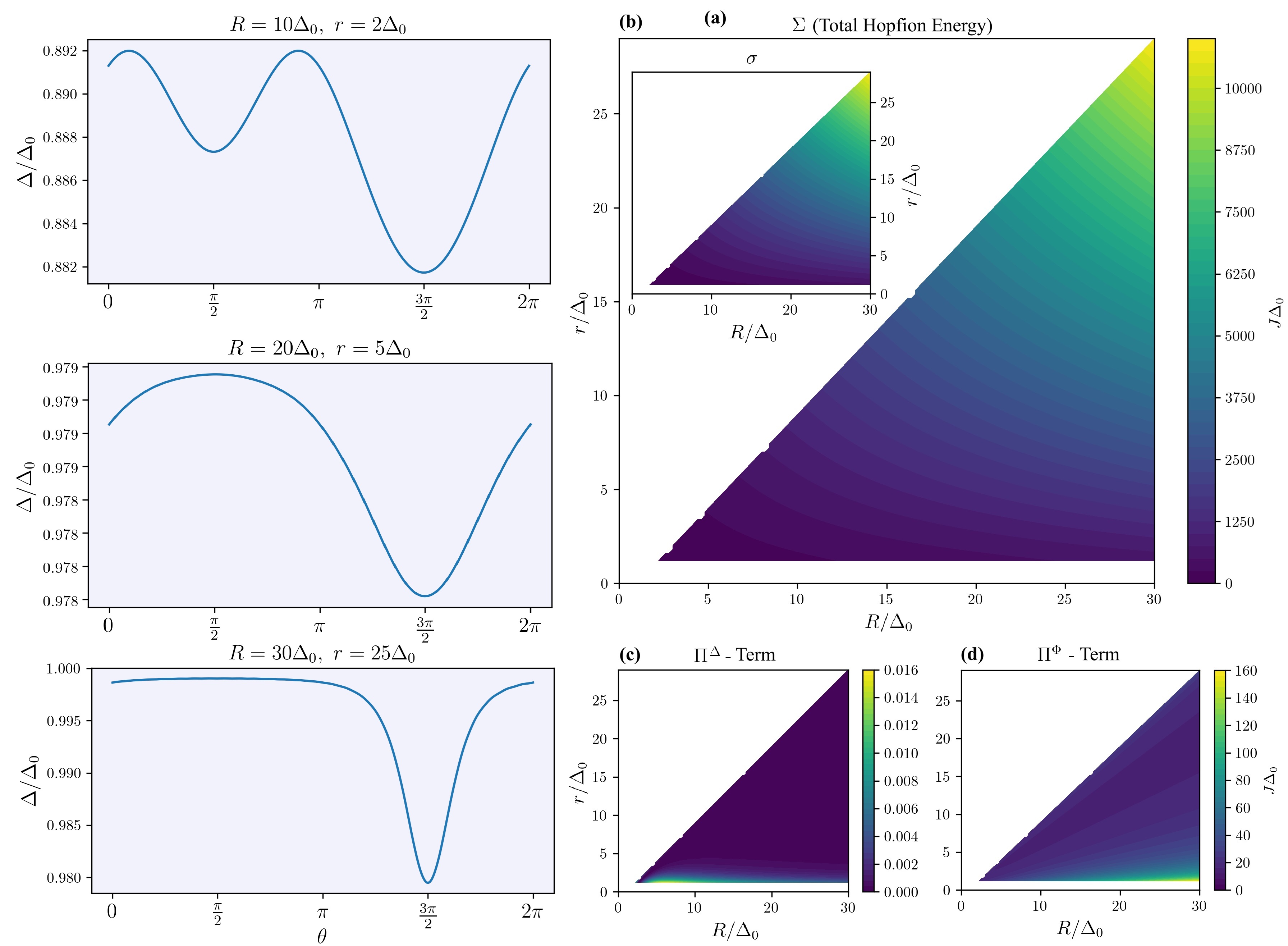} 
\caption{\label{fig:delta_sol} (a) Equilibrium solution for the domain-wall width $\Delta$ of a toroidal Hopfion for various values of the major radius $R$ and minor radius $r$. $\theta$ is the poloidal angle of the torus ranging over $0$ to $2\pi$. Note that the total variation of $\Delta$ over $\theta$ is less than a percent. (b) Total energy landscape, Eq.~(\ref{eq:sigmaDelta}), of a toroidal Hopfion over the major radius $R$ and minor radius $r$. The white regions indicate nonphysical parameters, such as when $r > R$ in the upper left which is no longer a torus or $r < \Delta$ at the bottom, where the torus is too thin. The apparent descent of the energy to smaller values of $r$ and $R$, and eventually to an unphysical value, indicates that given the ability to dissipate energy, the torus will shrink until it breaks the continuum limit. The inset shows the contribution to the energy from $\sigma$, the surface tension term. (c) Similar energy plot for the $\Pi^\Delta$ term, which contributes very little to the energy despite increasing by and order of magnitude for small Hopfions. (d) Likewise, the $\Pi^\Phi$ term has an overall small contribution to the total energy despite being roughly 3 orders of magnitude larger than $\Phi^\Delta$. All units of length are normalized by $\Delta_0$ and energy by $J\Delta_0$.}
\end{figure*}

\section{Example 2: Hopfion Decay Modes}\label{sec:five}
The next example we consider here concerns the decay modes of a toroidal Hopfion. At the time of submission, we are unaware of any standard micromagnetic model that hosts stable Hopfions. The issue appears to be, based on our own observations from micromagnetic simulations, that the competing energies will shrink the torus until the hole becomes only a few lattice spacings, at which point the continuum theory of topological invariants breaks and the discrete lattice collapses the center and eventually the entire Hopfion.

Here we model this process using the membrane theory. We construct the torus membrane with a major radius $R$, minor radius $r$ and an as yet unknown domain-wall thickness $\Delta$:
\begin{equation}\label{eq:torus}
    \mathbf{R} = 
\begin{pmatrix}
  (R+r\sin\theta)\cos\phi \\
  (R+r\sin\theta)\sin\phi \\
  r\cos\theta 
 \end{pmatrix}
\end{equation}
The torus membrane is thus parametrized by the two standard toroidal and poloidal angles $\phi$ and $\theta$ respectively, which both run from $0$ to $2\pi$ (see Fig. \ref{fig:torus}).

Next, for a standard Hopfion of winding 1, we use the simple form of the in-plane angle $\Phi = \theta$ which would also preserve energy and winding with a further global phase $\gamma$. Note that it does not have an explicit dependence on $\phi$ since $\Phi$ is defined locally with respect to a basis that is dependent on $\phi$ itself in Eq.~(\ref{eq:torusBasis}). We assume that due to the rotational symmetry of the torus (in $\phi$), the domain-wall thickness is only a function of the angle $\theta$. Thus the antisymmetric derivatives are
\begin{equation}
    \Delta^\alpha = 
\begin{pmatrix}
  0\\
  -\partial_\theta\Delta 
 \end{pmatrix},\text{ and }
    \Phi^\alpha = 
\begin{pmatrix}
  0\\
  -1 
 \end{pmatrix}.
\end{equation}

We can now test the local version of the Whitehead formula derived in Eq.~(\ref{eq:localHopf}) here. The local basis is simply
\begin{equation}\label{eq:torusBasis}
    \hat{\xi} = 
\begin{pmatrix}
  -\sin\phi \\
  \cos\phi \\
  0 
 \end{pmatrix} \text{ and }
 \hat{\chi} = \hat{\xi} \times\hat{z} = 
\begin{pmatrix}
  \cos\phi \\
  \sin\phi \\
  0 
 \end{pmatrix}
\end{equation}
making the Christoffel coefficients
\begin{align}
    \Gamma_{\chi\xi\theta} = 0 \text{ and } \Gamma_{\xi\chi\phi} = 1.
\end{align}
Substituting into the curvature determinant,
\begin{equation}
    K =
\begin{vmatrix}
    \partial_\theta\Phi & \partial_\phi\Phi \\
    0 & -1
\end{vmatrix}
= -\partial_\theta\Phi
\end{equation}
so that the local form of the invariant is
\begin{align}
    H =  \frac{-1}{(2\pi)^2}\int_0^{2\pi} d\theta \int_0^{2\pi} d\phi(-\partial_\theta\Phi) = 1.
\end{align}
This confirms the equation for the case of a standard Hopfion. We see that it essentially counts the two windings of $\Phi$ around both the toroidal and poloidal circles of the torus. One can generalize $\Phi$ to higher order Hopfions of winding $n$, by adopting a form $\Phi = n\theta$.

Carrying out the procedure of evaluating the relevant geometric quantities, we obtain the energy density of Eq.~(\ref{eq:EnDensity}):  
\begin{align}
    \Sigma = \sigma &+ \frac{J}{2}\Delta^i\Delta^j\Pi_{i,j}^\Delta + \frac{J}{2}\Phi^i\Phi^j\Pi^\Phi_{i,j} \\
    = \sigma &+ \frac{J}{2}(\partial_\theta\Delta)^2\Pi_{22}^\Delta + \frac{J}{2}(-1)^2\Pi_{22}^\Phi \\
    = \sigma &+ \frac{J\pi^2}{12}\frac{\Delta'^2}{g}\left(\frac{(R+r\sin\theta)^2}{\Delta} + \frac{7\pi^2}{20}\Delta\sin^2\theta\right) \nonumber\\
    &+\frac{J}{g}\left(\Delta (R+r\sin\theta)^2 + \frac{\pi^2}{12}\Delta^3\sin^2\theta\right), \label{eq:sigmaDelta}
\end{align}
which plays the role of a Lagrangian for the profile function $\Delta(\theta)$.  

The corresponding Euler–Lagrange equation is a nonlinear second-order ODE. While we do not attempt an analytic solution here, a representative numerical solution is plotted in Fig.~\ref{fig:delta_sol}(a) for a set of $R$ and $r$. A notable trend is that $\Delta(\theta)$ generally deviates only weakly from $\Delta_0$, but the variation becomes more pronounced for smaller Hopfions. In particular, as $R$ and $r$ approach the intrinsic length scale $\Delta_0$, the membrane thickness $\Delta$ is visibly reduced.  

To further explore possible decay channels, we examine the energy landscape as a function of the torus geometry, parametrized by its major and minor radii $R$ and $r$. The total energy is evaluated as  
\begin{equation}
    E[r,R] = \int d\theta\, d\phi \, r(R+r\sin\theta)\,\Sigma[r,R],
\end{equation}
under the assumption that the configuration $\Phi = \theta + \phi$ is preserved, and using the solution $\Delta(\theta)$ from Eq.~(\ref{eq:sigmaDelta}). The resulting energy landscape is shown in Fig.~\ref{fig:delta_sol}(b). The white regions correspond to unphysical parameter choices, such as the minor radius exceeding the major radius (top left) or shrinking below the domain-wall width (bottom). The overall trend demonstrates that the energy decreases monotonically with both $R$ and $r$, consistent with the surface tension contribution $\sigma$ scaling with the membrane area. This suggests that the true energy minimum lies within the excluded region where both radii collapse to the lattice scale, consistent with the physical observation that Hopfions stabilized solely by Heisenberg exchange and uniaxial anisotropy tend to shrink and eventually collapse under energy minimization.  

Figures~\ref{fig:delta_sol}(c) and (d) further show that the curvature contributions to the energy (the second and third terms in Eq.~(\ref{eq:sigmaDelta})) are typically orders of magnitude smaller than the surface tension term $\sigma$. However, these contributions grow significantly as $R$ and $r$ decrease, underscoring their importance in the behavior of small Hopfions.  

Possible decay paths can be identified by examining the energy dynamics in the presence of dissipation. Since the Hamiltonian has no explicit time dependence, the time derivative of the total energy is determined by the Landau-Lifshitz-Gilbert equation \cite{LANDAU1992,Gilbert2004}:  
\begin{equation}
    \partial_tE[R(t),r(t)] = -\frac{M_S\alpha}{\gamma}\int d^3r\,(\partial_t\mathbf{m})^2,
\end{equation}
where $M_S$ is the saturation magnetization, $\alpha$ is the Gilbert damping constant, and $\gamma$ is the gyromagnetic ratio. Because $\mathbf{m}$ is fully specified by the domain-wall ansatz through $\Delta(\theta)$, $\Phi(\theta,\phi)$, and the radii $R$ and $r$, the problem is analytically tractable. A detailed analysis of the dissipative dynamics along these decay paths will be an interesting direction for future work.  

This example illustrates how curvature influences energy and demonstrates the utility of the membrane formalism in capturing fine-grained information about the behavior of three-dimensional spin textures. It also provides a consistency check of our local form of the Whitehead formula in Eq.~(\ref{eq:localHopf}).

\section{Discussion and Conclusion}\label{sec:six}
Motivated by recent advances in the study of complex three-dimensional magnetic textures, we have developed an effective geometric framework for membranes in ferromagnets. By introducing a geometric ansatz for the membrane surface, we derived an effective energy functional from the full micromagnetic Hamiltonian that captures both curvature effects and the internal structure of the membrane. This formulation places magnetic domain walls in direct analogy with elastic surfaces, where curvature-dependent terms naturally play the role of surface tension and bending rigidity, and it is constructed in a coordinate-invariant manner to ensure independence from any particular parametrization. Within this framework, we established a local form of the Whitehead formula for the Hopf invariant, enabling a topological characterization directly in terms of membrane variables. As a concrete demonstration, we analyzed the case of a flat domain wall containing a vortex and further illustrated the framework with a Hopfion configuration where curvature plays a pivotal role and the local Whitehead formula reproduces the expected linking number.  

This membrane theory offers a versatile foundation for exploring the rich dynamics of a broad class of three-dimensional magnetic textures. It provides a natural framework for investigating topologically nontrivial structures such as torons, skyrmion tubes and strings, screw dislocations, and Hopfions, and beyond, each of which can be understood as specific membrane embeddings with distinct topological and geometric characteristics. Beyond static configurations, the formalism will accommodate dynamical phenomena, including the propagation of magnons along or through the membrane surface and the excitation of internal deformation modes such as breathing and flexural vibrations. The coupling between these collective modes and the geometry of the membrane opens a pathway for understanding how curvature and torsion mediate effective dynamics. In particular, the theory provides a natural setting in which to study the interplay between emergent electromagnetic fields and the geometry of the magnetization manifold, offering insights into how spin waves and textures interact with gauge fields arising from the magnetization dynamics.

To fully capture the behavior of realistic textures, especially in finite or confined geometries, a more complete treatment of boundary contributions is necessary. The variational principle applied to the effective energy must account for surface terms, which can lead to nontrivial boundary conditions that influence both the equilibrium configuration and the excitation spectrum. These boundary effects can be treated systematically within the membrane framework, and they contribute additional terms to the total energy functional, modifying Eq.~(\ref{eq:totalEnergy}). Depending on the physical context such as open surfaces, interfaces with different materials, or geometries with nontrivial topology, these boundary terms may encode important physics, including anchoring conditions, surface tension, or edge-bound modes. A careful examination of such contributions will be essential for extending the predictive power of the membrane theory to experimentally relevant systems.

A significant portion of the numerical calculations were conducted with Texas A\&M High Performance Research Computing resources.

\appendix
\section{Derivation of the Local Magnetization Ansatz for a Flat Membrane}\label{app:appA}

To construct the curved-membrane theory, we first determine the equilibrium structure of a flat domain-wall membrane. The key assumption is that, in the vicinity of any point on a curved membrane, the local magnetization profile along the wall normal is well approximated by that of a flat wall. This allows us to extract a universal “building block” profile that can be transplanted to curved geometries.

Consider an extended flat domain-wall, with a unit normal vector $\hat{n}$ that is not necessarily aligned with $\hat{z}$. The magnetization depends only on the coordinate $n$ along $\hat{n}$. Denote by $m_\parallel$ the component along $\hat{z}$ and $m_\perp$ the component in the plane perpendicular to $\hat{z}$, so that $m_\parallel^2 + m_\perp^2 = 1$. In this plane, $m_\perp$ makes an angle $\Phi$ with some fixed axis, which in general may depend on $n$.

From the micromagnetic Hamiltonian [Eq.~(\ref{eq:mmHamiltonian})], the energy per unit area $\sigma$ of the wall is
\begin{align}
\sigma = \frac{J}{2} \int \bigg[ (\partial_n m_\parallel)^2 & + (\partial_n m_\perp)^2 + m_\perp^2 (\partial_n \Phi)^2 \nonumber \\
& + \frac{2\lambda}{J} \left( 1 - m_\parallel^2 \right) \bigg] \, dn.
\end{align}
Minimization with respect to $\Phi$ yields $\Phi = \text{const}$, and variation with respect to $m_\parallel$ and $m_\perp$ gives the familiar Bloch-wall solution
\begin{align}
m_\parallel(n) = \tanh\left( \frac{n}{\Delta_0} \right), \quad
m_\perp(n) = \frac{1}{\cosh(n/\Delta_0)},
\end{align}
with the equilibrium wall thickness $\Delta_0 = \sqrt{J / 2\lambda}$.

In the flat-wall setting, $\Phi$ and $\Delta$ are constants. We now generalize by allowing the wall thickness $\Delta$ to deviate from $\Delta_0$, while retaining the same functional form of the profile:
\begin{align}    
m_\parallel(n) = \tanh(n/\Delta), \quad m_\perp(n) = \frac{1}{\cosh(n/\Delta)},
\end{align}
with $\Delta$ treated as a soft mode. The corresponding energy per unit area is
\begin{align}
\sigma(\Delta) = J \left( \frac{1}{\Delta} + \frac{\Delta}{\Delta_0^2} \right),
\end{align}
which expands near $\Delta_0$ as
\begin{align}
\sigma(\Delta) \approx \sigma_0 + \sigma_0 \frac{(\Delta - \Delta_0)^2}{\Delta_0^2}, \quad
\sigma_0 = \frac{2J}{\Delta_0}.
\end{align}
This expression plays the role of an effective surface tension, generalized to nonuniform $\Delta$.

To describe the spin orientation, it is convenient to introduce the orthonormal frame
\begin{align}\label{eq:globalBasis}
\hat{z}, \quad
\hat{\xi} = \frac{\hat{z} \times \hat{n}}{|\hat{z} \times \hat{n}|}, \quad
\hat{\chi} = \hat{\xi} \times \hat{z}.
\end{align}
At the wall center ($n=0$), the magnetization lies entirely in the $\hat{\xi}$–$\hat{\chi}$ plane, perpendicular to the uniform background along $\hat{z}$. At a distance $n$ from the wall center along $\hat{n}$, the magnetization components in this frame are
\begin{align}
m_z &= m_\parallel(n/\Delta), \nonumber \\
m_\xi &= m_\perp(n/\Delta) \sin\Phi, \quad
m_\chi = m_\perp(n/\Delta) \cos\Phi,
\end{align}
with $m_\parallel(0) = 0$ and $m_\parallel(|n/\Delta| \gg 1) \to \pm 1$.

Thus, in the vicinity of a flat domain-wall, the local magnetization ansatz takes the form
\begin{align}
\mathbf{m}(\mathbf{R} + n\hat{n}) = m_z \hat{z} + m_\xi \hat{\xi} + m_\chi \hat{\chi},
\end{align}
where $\mathbf{R}$ is the position on the wall center. This local representation forms the foundation for the curved-membrane formalism described in the main text and developed in the next Appendix.

\section{Energy Density for a Curved domain-wall Membrane}\label{app:appB}
Now we can construct the general result for a curved domain-wall membrane using the flat ansatz from Appendix A. Consider a membrane given parametrically by $\mathbf{R}(u, v)$ with the same orthonormal spin basis from Appendix A but now defined locally:
\begin{align}
\hat{n} = \frac{\partial_u \mathbf{R} \times \partial_v \mathbf{R}}{|\partial_u \mathbf{R} \times \partial_v \mathbf{R}|}, \quad
\hat{\xi} = \frac{\hat{z} \times \hat{n}}{|\hat{z} \times \hat{n}|}, \\
\hat{\chi} = \hat{\xi} \times \hat{z}.
\end{align}

Assuming the magnetization profile normal to the wall remains true to the flat case but with $\Delta$ defined locally on the membrane, $\Delta(u,v)$, we write the spin field using the same ansatz:
\begin{align}
&\mathbf{m}(\mathbf{R}) = m_\parallel\left(\frac{n}{\Delta(u,v)}\right) \hat{z} \nonumber\\
&+ m_\perp\left(\frac{n}{\Delta(u,v)}\right)
\left[\sin\Phi(u,v)\hat{\xi} + \cos\Phi(u,v)\hat{\chi} \right], \label{eq:ansatz}
\end{align}
where $n$ is again the distance along the normal $\hat{n}$ to the surface.

To compute the local form of the energy, we express the gradient in the Hamiltonian in terms of $\partial_{\mathbf{R}}$ through the chain rule. In the local coordinates this is
\begin{align}
\partial_{\mathbf{R}} &= \frac{\mathbf{A}_v \times (\mathbf{A}_u \times \mathbf{A}_v)}{|\mathbf{A}_u \times \mathbf{A}_v|^2} \partial_u +
\frac{\mathbf{A}_u \times (\mathbf{A}_v \times \mathbf{A}_u)}{|\mathbf{A}_u \times \mathbf{A}_v|^2} \partial_v + \hat{n} \, \partial_n,
\end{align}
where $\mathbf{A}_u = \partial_u \mathbf{R} + n \partial_u \hat{n}$ and similarly for $\mathbf{A}_v $.

The volume element now becomes
\begin{align}
dV = \hat{n} \cdot (\mathbf{A}_u \times \mathbf{A}_v) \, du \, dv \, dn,
\end{align}
and the cross product $\mathbf{A}_u \times \mathbf{A}_v$ can be expanded in powers of the normal distance $n$ as
\begin{align}
\mathbf{A}_u \times \mathbf{A}_v = \hat{n} \sqrt{\det g} \left(1 - n \, \text{Tr}\, S + n^2 \det S + \cdots \right),
\end{align}
where 
\begin{equation}
  g_{\alpha\beta} = \partial_\alpha \mathbf{R} \cdot \partial_\beta \mathbf{R}  
\end{equation}
is the induced metric and
\begin{equation}
 S_{\alpha\beta} = \hat{n} \cdot \partial^2_{\alpha\beta} \mathbf{R}   
\end{equation}
is called the shape operator (also known as the second fundamental form \cite{doCarmo1976}). Since it takes second derivatives of $\mathbf{R}$ it is tied intimately to curvature and naturally vanishes for flat membranes.

In the large curvature limit, where the wall width $\Delta$ is small compared to the radius of curvature, higher-order terms in $n$ can be neglected:
\begin{align}
\mathbf{A}_u \times \mathbf{A}_v \approx \hat{n} \sqrt{\det g}.
\end{align}

Substituting into the energy functional, carrying out the derivatives, and integrating over $n$, the total energy becomes
\begin{align}
\mathcal{H}[\Delta, \Phi, \mathbf{R}] = \int du \, dv \, \sqrt{\det g} \, \Sigma(\Delta, \Phi, \mathbf{R}),
\end{align}
where
\begin{align}
\Sigma = \sigma(\Delta) + \frac{J}{2} \left( \Delta_\alpha \Delta_\beta \Pi^{\Delta}_{\alpha\beta} + \Phi_\alpha \Phi_\beta \Pi^{\Phi}_{\alpha\beta} \right),
\end{align}
with
\begin{align}
\Delta_\alpha = \epsilon_{\alpha\beta} \partial_\beta \Delta, \text{ and }
\Phi_\alpha = \epsilon_{\alpha\beta} \partial_\beta \Phi.
\end{align}
The geometric coupling tensors are given by
\begin{align}
    \Pi^\Phi_{\alpha,\beta} = \frac{2\Delta}{g}(g_{\alpha,\beta} + \frac{\pi^2}{12}\Delta^2S_\alpha^\gamma S_{\gamma,\beta}) \\
    \Pi^\Delta_{\alpha,\beta} = \frac{\pi^2}{6\Delta g}(g_{\alpha,\beta} + \frac{7\pi^2}{20}\Delta^2S_\alpha^\gamma S_{\gamma,\beta})
\end{align}
and capture the leading curvature corrections to the domain-wall energy as well as reduce to the flat-wall case when the curvature vanishes ($ S_{\alpha\beta} = 0 $).

The energy functional above is invariant under reparametrizations of the membrane coordinates $(u,v)$ and can be expressed in a covariant, coordinate-free form. The transverse derivatives $\Delta_\alpha = \epsilon_{\alpha\beta}\partial_\beta \Delta$ and $\Phi_\alpha = \epsilon_{\alpha\beta}\partial_\beta \Phi$ are the Hodge duals of the exterior derivatives of the scalar fields $\Delta:\mathbf{R}\to\mathbb{R}$ and $\Phi:\mathbf{R}\to\mathbb{S}^1$ on the Riemannian surface $\mathbf{R}\subset\mathbb{R}^3$ with induced metric $g_{\alpha\beta}$ \cite{Nakahara2003}.

In formal language, the energy density is
\begin{align}
\Sigma = \sigma(\Delta) + \frac{J}{2}\left( \langle d\Delta, \Pi^\Delta[d\Delta] \rangle_g + \langle d\Phi, \Pi^\Phi[d\Phi] \rangle_g \right),
\end{align}
where $\langle\cdot,\cdot\rangle_g$ is the metric inner product, and $\Pi^\Delta$, $\Pi^\Phi$ are symmetric $(1,1)$-tensors obtained from transverse-coordinate integrals involving the shape operator $S$ (the second fundamental form). 

In the flat limit, $\Pi^\Delta$ and $\Pi^\Phi$ reduce to operators proportional to $g^{\alpha\beta}$, while curvature introduces anisotropic corrections via contractions of $S$ with itself. This provides a geometric stiffness tensor whose form is dictated by both intrinsic and extrinsic curvature. The result places the problem within geometric field theory on submanifolds, where scalar modes couple covariantly to membrane shape and topology.

\section{Hopf Invariant in Local Coordinates}\label{app:appC}
The emergent gauge field $\mathbf{A}$ can be written in a spherical parametrization through the Berry connection as
\begin{align}
\mathbf{A} = \frac{1}{2}(1-\cos\Theta)\nabla \Phi,
\end{align}
and together with the corresponding magnetic field $\mathbf{F} = \nabla \times \mathbf{A}$ defines the Hopf invariant:
\begin{align}
H = -\frac{1}{(2\pi)^2} \int d^3x \, \mathbf{F} \cdot \mathbf{A}.
\end{align}

It is vital to recognize that the Berry connection $\mathbf{A}$ is not globally defined \cite{Ryder1980}. In fact,
\begin{align}
\mathbf{A} = \frac{1}{2}(1 - \cos\Theta) \nabla \Phi
\end{align}
is regular near $\Theta = 0$, but becomes singular at $\Theta = \pi$, and vice versa for the alternative choice of sign
\begin{align}
\mathbf{A} = \frac{1}{2}(1 + \cos\Theta) \nabla \Phi
\end{align}
Therefore, a single global gauge cannot cover the whole space. Denote these two patches as $\mathbf{A}_\uparrow$ and $\mathbf{A}_\downarrow$ respectively, and note that they are equal on the membrane where $\cos\Theta = 0$.

To evaluate the invariant, we first resolve the derivatives of the magnetization field in the local coordinate basis, namely $\partial_n\mathbf{m},~\partial_u\mathbf{m}$, and $\partial_v\mathbf{m}$. These derivatives are then used to construct the emergent field components 
\begin{equation}
    F_i = \frac{1}{2}\epsilon_{ijk}\mathbf{m}\cdot(\partial_j\mathbf{m}\times\partial_k\mathbf{m}).
\end{equation}
 Because the texture is localized near the membrane, we integrate over the normal direction $n$ analytically. To properly account for the gauge patches, the domain is split into two neighborhoods, one covering the region outside the membrane (where $\Theta \to 0$) and one for the interior (where $\Theta \to \pi$) and use $\mathbf{A}_\uparrow$ and $\mathbf{A}_\downarrow$ respectively.

Performing the integration over $n$, this procedure reduces the invariant to a purely two-dimensional surface integral:
\begin{align}
H = -\frac{1}{(2\pi)^2} \int du \, dv \, \mathcal{K}(u, v),
\end{align}
with integrand
\begin{align}
\mathcal{K}(u,v) = \det \begin{pmatrix}
\partial_u \Phi & \partial_v \Phi \\
\Gamma_{\chi\xi u} & \Gamma_{\chi\xi v}
\end{pmatrix}.
\end{align}

The coefficients $\Gamma_{\chi\xi u} = \partial_u \hat{\xi}\cdot \hat{\chi}$ and $\Gamma_{\chi\xi v} = \partial_v \hat{\xi}\cdot \hat{\chi}$ describe the rotation of the local in-plane frame $(\hat{\xi},\hat{\chi})$ along the membrane coordinates. In this language, the in-plane spin angle $\Phi$ behaves as a local $U(1)$ phase field, while $\Gamma_{\chi\xi}$ plays the role of a connection one-form. The Hopf density can then be expressed as the 2-form
\begin{align}
\mathcal{K} = d\Phi \wedge \omega, \qquad \omega(\partial_\alpha)=\Gamma_{\chi\xi\alpha},
\end{align}
which captures the coupling between the in-plane angle $\Phi$ and the geometry of the membrane frame. This expression highlights that the Hopf charge arises from the interplay of spin twisting and membrane curvature, and geometrically corresponds to the linking of spin preimages in three-dimensional space. In compact form over the surface $\Omega$, the invariant reads
\begin{align}
H = -\frac{1}{(2\pi)^2}\int_\Omega d\Phi\wedge \omega,
\end{align}
showing explicitly how the Hopf charge can be evaluated from surface data alone.

\bibliographystyle{apsrev4-2}
\bibliography{membrane_statics}
\end{document}